\RequirePackage{fix-cm}
\documentclass[epjc3]{svjour3}       % onecolumn (second format)
\def\makeheadbox{{\vspace*{-4mm}\flushright
LU-TP 12-44\\
MCnet 12-16\\
arXiv:yymm.nnnn [hep-ph]\\~
}}
\usepackage{graphicx}
\usepackage{epsfig}
\usepackage{xspace}
\usepackage[utf8]{inputenc}
\usepackage{cite}
\usepackage{epsfig}
\usepackage{xspace}
\usepackage{graphics}
\usepackage{inputenc}

\def\hide#1{}

\newcommand{\pythia}{P\protect\scalebox{0.8}{YTHIA}\xspace}

\newcommand{\pytppp}{P\protect\scalebox{0.8}{YTHIA}8\xspace}

\newcommand{\as}{\ensuremath{\alpha_{\mathrm{s}}}}

\newcommand{\aem}{\ensuremath{\alpha_{\mathrm{EM}}}}

\newcommand{\particle}[1]{\ensuremath{\mathrm{#1}}}
\newcommand{\antiparticle}[1]{\ensuremath{\bar{\mathrm{#1}}}}
\newcommand{\el}{\particle{e}}

\newcommand{\q}{\particle{q}}

\newcommand{\qbar}{\antiparticle{q}}

\newcommand{\ee}{\ensuremath{\el^+\el^-}}
\newcommand{\tee}{\ensuremath{\el^+\el^-}\xspace}

\newcommand{\done}[1]{}

\def\mrm#1{\mathrm{#1}}
\newcommand{\ie}{i.e.\xspace}
\newcommand{\eg}{e.g.\xspace}

\def\sub#1{\ensuremath{_{\mrm{#1}}}}

\def\f2d3{\ensuremath{F_2^{\mrm{D}3}}}

\providecommand{\eqref}[1]{eq.~(\ref{#1})\xspace}
\renewcommand{\eqref}[1]{eq.~(\ref{#1})\xspace}

\newcounter{aenumct}

\newcounter{ienumct}

{\end{list}}

\newcounter{enumct}

\begin{document}

\title{Fooling Around with the Sudakov Veto Algorithm\thanks{Work supported in parts by
    the Swedish research council (contracts 621-2009-4076 and
    621-2010-3326).}%\thanks{Grants or other notes
}


\author{Leif Lönnblad\thanksref{inst:lund}}


\institute{Department of Astronomy and Theoretical Physics, Lund
  University, Sweden.\label{inst:lund}}

\date{}

\maketitle

\begin{abstract}
  The Sudakov veto algorithm for generating emission and no-emission
  probabilities in parton showers is revisited and some reweighting
  techniques are suggested to improve statistics by oversampling in
  specific cases.

\end{abstract}

\section{Introduction}
\label{intro}

Sudakov \cite{Sudakov:1954sw} form factors or no-emission probabilities, are used in all
parton shower (PS) programs, to ensure exclusive final states and, at
the same time, to resum leading (and sub-leading) logarithmic virtual
contributions to all orders. The way these enter into the shower is
through the ordering of emissions, typically in transverse momentum or
angle. In the standard case we have an inclusive splitting probability
of a parton $i$ into partons $j$ and $k$ given by
\begin{equation}
  d{\cal P}_{i,jk}(t,z)=\frac{\as}{2\pi}P_{i,jk}(z)dtdz,
  \label{eq:split}  
\end{equation}
where $t=\log{p_\perp^2/\Lambda_{\mbox{\scriptsize QCD}}^2}$ is the
ordering variable, $z$ represents the energy sharing between $j$ and
$k$, and where we have integrated over azimuth angle in the
Altarelli--Parisi splitting function $P_{i,jk}(z)$. Starting from some
maximum scale $t_0$ we then want to find the exclusive probability of
the first emission, which we get from the inclusive splitting
probability by multiplying with the probability that there is no
emission before the first emission,
\begin{equation}
  d{\cal P}^{\mbox{\scriptsize first}}_{i,jk}(t,z)=
  \frac{\as}{2\pi}P_{i,jk}(z)dtdz\times\Delta(t_0,t).
  \label{eq:firstsplit}  
\end{equation}
Here, $\Delta(t_0,t)$ is this no-emission probability, or the Sudakov
form factor, given by
\begin{equation}
  \Delta(t_0,t)=
  \exp\left(-\int_t^{t_0}dt'dz \frac{\as}{2\pi}P_{i,jk}(z)\right).
  \label{eq:basicsud}
\end{equation}

In principle the Sudakov form factor can be calculated analytically.
However, often the integration region in the $z$-integral can be
non-trivial, and most PS programs today prefer to calculate it
numerically using the so-called Sudakov veto algorithm
\cite{Sjostrand:2006za}. The trick here is to find a simple function
which is everywhere larger than $P_{i,jk}$ and which is easy to
integrate, and by systematically vetoing emissions generated according
to this overestimated function, a correct no-emission probability is
obtained.

The Sudakov veto algorithm (SVA) is normally used for purely probabilistic
processes, but recently it has been generalized to also be used in
cases where the function being exponentiated is not positive definite
\cite{Platzer:2011dq,Hoeche:2011fd}.

In this article we shall investigate other modifications of the
Sudakov veto algorithm, where we try to increase the statistical
precision in some special cases by oversampling techniques, but we
will also briefly discuss the issue of negative contributions to
splitting functions.

In section \ref{sec:rewe-ckkw-like} we will investigate the usage of
the SVA in CKKW-like \cite{Catani:2001cc,Lonnblad:2001iq} algorithms,
where parton showers are matched with fixed order inclusive matrix
elements (MEs). Here, the SVA is used to make the inclusive MEs
exclusive by multiply them with no-emission probabilities taken from a
parton shower. One problem with this procedure is that every event
prduced with the matrix element generator is either given the weight
zero or one, which becomes very inefficient if the cutoff used in the
ME-generation is small. We will find that by introducing oversampling,
a weight can be calculated which is never zero, but nevertheless will
give the correct no-emission probability. In section
\ref{sec:rewe-ckkw-like} we will also discuss an extension of the CKKW
algorithm to include next-to-leading order (NLO) MEs
\cite{Lavesson:2008ah} where the SVA is used to extract fixed orders
of $\as$ form the parton shower to avoid double counting of
corresponding terms in the NLO calculation.

Then, in section \ref{sec:rewe-comp-proc} we will consider cases where
a parton shower includes different competing processes, where some of
them are very unlikely. This is the case in \eg the \pythia\ parton
shower, where photon emissions off quarks are included together with
standard QCD splittings. Since $\aem$ is much smaller than $\as$ it is
very time consuming to produce enough events containing hard photons
to get reasonable statistics. We shall see that a naive oversampling
of the photon emissions has unwanted effects on the total no-emission
probability, and that a slightly more involved procedure is
needed. The method presented is different from the one introduced by
Höche et al.\ in \cite{Hoeche:2009xc}, but is equally valid. It turns
out that both these methods can be used to include negative terms in
the splitting functions.

But first we shall revisit the derivation of the SVA, as we will use
many of the steps from there when we investigate the different
oversampling techniques.

\section{The Sudakov Veto Algorithm}
\label{sec:sudak-veto-algor}

Here we follow the derivation found in \cite{Sjostrand:2006za} and
\cite{Buckley:2011ms}. Although we normally have competing processes,
we will first simplify the notation by just considering one possible
splitting function, which we will integrate over $z$,
\begin{equation}
  \Gamma(t)=\int_{z_{\min}(t)}^{z_{\max}(t)}\frac{\as(t)}{2\pi}P(z)dz.
\end{equation}
The no-emission probability simply becomes
\begin{equation}
  \Delta(t_0,t_c)=e^{-\int_{t_c}^{t_0}\Gamma(t)dt}.
\end{equation}

If $\Gamma(t)$ can be integrated analytically, and if the primitive
function, $\check{\Gamma}$ has a simple inverse, it is easy to see
that we can generate the $t$-value of the first emission by simply
generate a random number, $R$, between zero and unity and obtain
\begin{eqnarray}
  && \int_t^{t_0}dt'\Gamma(t')\Delta(t_0,t')=
  R\int_0^{t_0}dt'\Gamma(t')\Delta(t_0,t')\nonumber\\
  \Rightarrow && 1-\Delta(t_0,t)=R\left(1-\Delta(t_0,0)\right)\nonumber\\
  && 1-\exp{(\check{\Gamma}(t)-\check{\Gamma}(t_0))}=R\nonumber\\
  \Rightarrow && t =
  \check{\Gamma}^{-1}\left(\check{\Gamma}(t_0)+\log(1-R)\right),
  \label{eq:anagen}
\end{eqnarray}
where we have assumed that $\Gamma(t)$ is divergent for small $t$,
such that $\Delta(t,0)=0$, an assumption we will come back to below.

Now, in most cases the integration of $\Gamma$ is not possible to do
analytically, and if it is, the inverse function may be
non-trivial. This is the case which is solved by the SVA. All we need
to do is to find a nicer function, $\hat{\Gamma}$, with an analytic
primitive function, which in turn has simple inverse, such that it
everywhere overestimates $\Gamma$,
\begin{equation}
  \hat{\Gamma}(t)\ge\Gamma(t),\quad\forall t.
\end{equation}
With this function we can construct a new no-emission probability
\begin{equation}
  \hat{\Delta}(t_0,t_c)=e^{-\int_{t_c}^{t_0}\hat{\Gamma}(t)dt}  
\end{equation}
which is everywhere an \textit{under}estimate of $\Delta(t_0,t_c)$,
and we can generate the first $t$ according to it. As in the standard
\textit{accept--reject} method, we now accept the generated value with
a probability $\Gamma(t)/\hat{\Gamma}(t)<1$. However, contrary to the
standard method, if we reject the emission, we replace $t_0$ with the
rejected $t$-value before we generate a new $t$. Loosely speaking, we
have underestimated the probability that the emission was not made
above $t$, so we need not consider that region again. We now continue
generating downwards in $t$ until we either accept a $t$-value, or
until the generated $t$ drops below $t_c$ at which point we give up
and conclude that there was no emission above $t_c$.

To see how this works more precisely, we look at the total probability
of not having an emission above $t_c$,
\begin{equation}
  {\cal P}\sub{tot}=\sum_{n=0}^\infty{\cal P}_n,
\end{equation}
where ${\cal P}_n$ is the probability that we have rejected $n$
intermediate $t$-values. To start with, we have
\begin{eqnarray}
  {\cal P}_0&=&\hat{\Delta}(t_0,t_c)\nonumber\\
  {\cal P}_1&=&\int_{t_c}^{t_0}dt\hat{\Gamma}(t)\hat{\Delta}(t_0,t)
  \left[1-\frac{\Gamma(t)}{\hat{\Gamma}(t)}\right]\hat{\Delta}(t,t_c)\nonumber\\
  &=&\hat{\Delta}(t_0,t_c)\int_{t_c}^{t_0}dt
  \left[\hat{\Gamma}(t)-\Gamma(t)\right],
\end{eqnarray}
where for ${\cal P}_1$ we first have the probability that we generate
a value $t$ and then throw it away with probability
$1-\Gamma(t)/\hat{\Gamma}(t)$ and then the probability that we do not
generate anything below $t$. Similarly we get
\begin{eqnarray}
  {\cal P}_2&=&\int_{t_c}^{t_0}dt_1
  \hat{\Gamma}(t_1)\hat{\Delta}(t_0,t_1)
  \left[1-\frac{\Gamma(t_1)}{\hat{\Gamma}(t_1)}\right]\nonumber\\
  &&\qquad\times\int^{t_1}_{t_c}dt_2\hat{\Gamma}(t_2)\hat{\Delta}(t_1,t_2)
  \left[1-\frac{\Gamma(t_2)}{\hat{\Gamma}(t_2)}\right]\hat{\Delta}(t_2,t_c)
  \nonumber\\
  &=&\hat{\Delta}(t_0,t_c)\frac{1}{2}\left(\int_{t_c}^{t_0}dt
  \left[\hat{\Gamma}(t)-\Gamma(t)\right]\right),
\end{eqnarray}
from noting that
$\Delta(t_0,t_1)\Delta(t_1,t_2)\Delta(t_2,t_c)=\Delta(t_0,t_c)$ and
that the nested integral can be easily factorized.

This is easily generalized to an arbitrary number of vetoed emissions and we get
\begin{eqnarray}
  \sum_{n=0}^{\infty}{\cal P}_n&=&\sum_{n=0}^{\infty}\hat\Delta(t_0,t_c)
  \frac{1}{n!}\left(\int_{t_c}^{t_0}dt
    \left[\hat{\Gamma}(t)-\Gamma(t)\right]\right)^n\nonumber\\
  &=&\hat\Delta(t_0,t_c)e^{\int_{t_c}^{t_0}dt\left[\hat{\Gamma}(t_1)-\Gamma(t)\right]}\nonumber\\
  &=&\Delta(t_0,t_c),
\end{eqnarray}
which is the no-emission probability we want.

Here we have ignored the additional variables involved in the
emissions. It is easy to see that if we have a simple overestimate of
the splitting function
\begin{equation}
  \hat{P}(t,z)\ge\frac{\as(t)}{2\pi}P(z),\quad \forall t,z
\end{equation}
we can construct our $\hat{\Gamma}$ as
\begin{equation}
  \hat{\Gamma}(t)=\int_{\hat{z}_{\min}(t)}^{\hat{z}_{\max}(t)}\hat{P}(t,z)dz
\end{equation}
with $\hat{z}_{\max}(t)\ge z_{\max}(t)$ and $\hat{z}_{\min}(t)\le
z_{\min}(t)$ overestimates the integration region in $z$. For each $t$
we generate, we then also generate a $z$ in the interval
$[\hat{z}_{\min}(t),\hat{z}_{\max}(t)]$ according to the probability
distribution
\begin{equation}
  {\cal P}(z) = \frac{\hat{P}(t,z)}{\hat{\Gamma}(t)}
\end{equation}
and we then only keep the emission with the probability
\begin{equation}
  \frac{\frac{\as(t)}{2\pi}P(z)\Theta(z-z_{\min}(t)\Theta(z_{\max}(t)-z))}
  {\hat{P}(t,z)}.
\end{equation}
Although the formulae become more cluttered, it is straight-forward to
show, by going through the steps above that this will give the correct
distributions of emissions.

If we now go back to \eqref{eq:anagen}, we there assumed that $\Gamma$
diverges at zero such that $\Delta(t,0)=0$. This is, of course, not
necessarily the case, as pointed out in
\cite{Platzer:2011dq}. However, here we will only concern ourselves
with emissions above some cutoff $t_c>0$, and we can always add to our
overestimate a term which is zero above $t_c$ but which diverges at
$t=0$. The fact that nowhere in the veto algorithm does anything thing
depend on the form of $\hat{\Gamma}$ below $t_c$, means that we do not
even need to specify how it diverges, it is enough to assume that it
does.

In parton showers one always have many different emission
probabilities. We typically have several possible emissions for each
parton. The SVA is easily extended to this situation by noting that
the no-emission probability factorizes for the sum of different
splittings,
\begin{equation}
  \Delta(t_0,t_c)=e^{-\int_{t_c}^{t_0}\sum_a\Gamma_a(t)dt}=
  \prod_a\Delta_a(t_0,t_c).
\end{equation}
The first emission is then generated with a $t$-value given according to 
\begin{equation}
  d{\cal P}=\sum_a\Gamma_a(t)\Delta(t_0,t)dt,
\end{equation}
and then randomly selecting one of the processes with weight
\begin{equation}
  \frac{\Gamma_a(t)}{\sum_b\Gamma_b(t)}.
\end{equation}
It is easily shown that generating one $t$ for each possible splitting
according to
\begin{equation}
  d{\cal P}_a=\Gamma_a(t_a)\Delta_a(t_0,t_a)dt_a,
\end{equation}
and then selecting the splitting which gave the largest $t_a$, gives
the same result.

\section{Reweighting in CKKW-like procedures}
\label{sec:rewe-ckkw-like}

In CKKW we want to multiply partonic states generated by a ME
generator with Sudakov form factors. We take the partonic state and
project it onto a parton-shower history. An $n$-parton state is then
reconstructed as a series of parton shower emissions with emission
scales $\{t_1,\ldots,t_n\}$ and the corresponding intermediate states
$\{S_0,\ldots,S_n\}$ where $S_n$ is the one generated by the ME.
We then want to multiply by the no-emission factors
\begin{equation}
  \Delta_i(t_i,t_{i+1})=e^{-\int_{t_{i+1}}^{t_i}dt\Gamma_i(t)}
\end{equation}
where $\Gamma_i$ is the sum of the splitting functions from the
partons in state $S_i$.

What we can do is to simply put the state $S_i$ into the parton shower
program and ask it to generate one emission starting from the scale
$t_i$. If the generated emission has a scale $t>t_{i+1}$ we throw the
whole partonic event away and ask the ME generator to produce a new
state. The probability for this not to happen is exactly
$\Delta_i(t_i,t_{i+1})$ and the procedure corresponds to reweighting
the ME state with the desired no-emission probability.

The problem is that if the ME state corresponds to very low scales, we
will throw away very many events, which is very inefficient and may
result in poor statistics.

A way to improve this situation is to introduce a boost factor for the
splittings $\Gamma_i\to\tilde\Gamma_i=C\Gamma_i$, with $C>1$, and
multiply the overestimate $\hat\Gamma_i$ with the same factor. As
before this just gives a simple overestimate of the splitting
function, which we know how to handle from section
\ref{sec:sudak-veto-algor}. But rather than throwing an emission away
with an extra probability $1/C$ (and not veto the event) we can always
reject the emission but multiply the whole event with a weight
$1-1/C$. The total weight of the event will then be the sum of all
possible ways we can veto a generated emission (we here assume that the
normal rejection procedure has already been applied)
\begin{eqnarray}
  \langle w_i\rangle&=&
  \sum_{n=0}^{\infty} (1-1/C)^n\tilde\Delta_i(t_i,t_{i+1})\frac{1}{n!}
  \left(\int_{t_{i+1}}^{t_i}dt\tilde\Gamma_i(t)\right)^n\nonumber\\
  &=&\sum_{n=0}^{\infty} \tilde\Delta_i(t_i,t_{i+1})\frac{1}{n!}
  \left(\int_{t_{i+1}}^{t_i}dt
    \left[\Gamma_i(t)-\tilde\Gamma_i(t)\right]\right)^n\nonumber\\
  &=&\Delta_i(t_i,t_{i+1})
\end{eqnarray}
In this way we get the right weight but we never throw away an event.

In the NLO version of the CKKW-L \cite{Lavesson:2008ah} we also want
to calculate the integral of the splitting function,
$\int_{t_{i+1}}^{t_i}dt \Gamma(t)_i$ which is used as a way of
subtracting the fixed first orders result from the exponentiation and
then replace it with the correct NLO result. The way this was done in
\cite{Lavesson:2008ah} was similar to the procedure above. The shower
is started, and each emission is vetoed, but the number of emissions
above $t_{i+1}$ was counted and it was noted that the average number
of vetoed emissions is given by
\begin{eqnarray}
  \langle n\rangle&=&\sum_{n=0}^{\infty} n\Delta_i(t_i,t_{i+1})
  \frac{1}{n!}\left(\int_{t_{i+1}}^{t_i}dt
    \Gamma(t)_i\right)^n\nonumber\\
  &=&\int_{t_{i+1}}^{t_i}dt\Gamma_i(t) \times\sum_{n=1}^{\infty}
  \Delta_i(t_i,t_{i+1})\frac{1}{(n-1)!}\left(\int_{t_{i+1}}^{t_i}dt
    \Gamma_i(t)\right)^{n-1}\nonumber\\
  &=&\int_{t_{i+1}}^{t_i}dt
  \Gamma_i(t)
\end{eqnarray}

To factor out other fixed order terms we note also that
\begin{equation}
  \langle n(n-1)\cdots (n-m)\rangle=\frac{1}{m!}\left(\int_{t_{i+1}}^{t_i}dt
    \Gamma_i(t)\right)^m,
\end{equation}
which means we can pick out higher-order terms in $\as$ used in the
shower.

Again, the statistics can become a bit poor if most events yield the
weight zero (which is the case if for large merging scales when the
no-emission probability is close to unity), and only a few have
non-zero values. We can instead again introduce the boost factor, $C$,
and rather than simply counting the number of emission we take the
weight $n/C$.

We note that in general this $C$ need not be a simple constant, it can
be a function of the scale (or any other variable in the splitting.)
This is used in the NLO version of CKKW-L, where the leading order
$\as$ term in the expansion of the no-emission probability is needed
at a fixed renormalization scale, $\mu_R$, while in the shower we have
a coupling running with the transverse momentum. Therefore, rather
than counting the the number of emissions, we sum up ratios of fixed
and running $\as$ for the emissions which are generated and
discarded. Introducing
\begin{equation}
  \Gamma_R(t)=\int_{z_{\min}(t)}^{z_{\max}(t)}\frac{\as(\mu_R)}{2\pi}P(z)dz=
  \frac{\as(\mu_R)}{\as(t)}\Gamma(t)
\end{equation}
We then get on the average a weight
\begin{eqnarray}
  \langle w\rangle&=&
  \sum_{n=0}^{\infty}\Delta(t_0,t_c)\int_{t_c}^{t_0}dt_1\Gamma(t_1)
  \cdots\int_{t_c}^{t_{n-1}}dt_n\Gamma(t_n)
  \left(\sum_{i=1}^n\frac{\as(\mu_R)}{\as(t_i)}\right)\nonumber\\
 &=& \int_{t_c}^{t_0}dt\Gamma_{R}(t)\times
 \sum_{n=1}^{\infty}\Delta(t_0,t_c)\frac{1}{(n-1)!}
 \left(\int_{t_c}^{t_0}dt\Gamma(t)\right)^{n-1}\nonumber\\
 &=& \int_{t_c}^{t_0}dt\Gamma_{R}(t),\label{eq:fixalp}
\end{eqnarray}
by working a bit on the symmetrized nested integrals of different
functions and we get what we desired. To obtain higher powers of the
integral with fixed \as, it is easy to show that \eg the average sum
of triplets
\begin{equation}
  \sum_{i\ne j\ne k}\frac{\as(\mu_R)}{\as(t_i)}
  \frac{\as(\mu_R)}{\as(t_j)}\frac{\as(\mu_R)}{\as(t_k)}
\end{equation}
will give
\begin{equation}
  \frac{1}{3!}\left(\int_{t_c}^{t_0}dt\Gamma_{R}(t)\right)^3.
\end{equation}

We also note that for initial-state splittings, the integral over
splitting functions also contain ratios of parton density functions,
\begin{equation}
  \frac{f_b(x/z,t)}{f_a(x,t)},
\end{equation}
where $a$ is the incoming parton before, and $b$ is the parton after
the splitting. What is needed in the NLO-version of CKKW-L in this
case is the integral for a given factorization scale, which is
obtained by simply changing the \as-weight in \eqref{eq:fixalp} to
\begin{equation}
  \frac{\as(\mu_R)}{\as(t)}
  \frac{f_b(x/z,\mu_F)}{f_b(x/z,t)}\frac{f_a(x,t)}{f_a(x,\mu_F)},
\end{equation}
where $z$ is the energy fraction of the vetoed generated splitting.
The derivation in \eqref{eq:fixalp} becomes a bit more cumbersome, but
is straight forward.

\section{Reweighting competing processes}
\label{sec:rewe-comp-proc}

Often we have many different competing splitting processes. The
example we shall use here is the process of a quark radiating a gluon
($\Gamma_g$) competing with the process of the same quark radiating a
photon ($\Gamma_\gamma$). Since generating the latter much less likely
because of the smallness of $\aem$ as compared to $\as$, the
generation may become very inefficient if we are interested in
observables related to an emitted photon.

In principle we could again consider introducing a boost factor $C>1$
and replace $\Gamma_\gamma(t)$ with $\tilde\Gamma_\gamma(t) =
C\Gamma_\gamma(t)$ and do the same with the overestimate
$\hat\Gamma_\gamma$. As long as $\tilde\Gamma_\gamma(t) \ll
\Gamma_g(t)$ we can reweight each event containing $n$ photons with a
factor $1/C^n$ and get approximately the correct results for the
observables. However this only gives the right emission probability,
not the correct no-emission probability.

Instead we adopt different procedure. Every time we generate a photon
emission (accepted with probability
$\Gamma_\gamma/\hat\Gamma_\gamma$), we veto it anyway with a
probability 0.5. If we veto it, we also reweight the whole event with
a factor $2-2/C$, while if we keep it, we reweight the whole event
with a factor $2/C$. Clearly the emissions will still be correctly
weighted, $0.5\times2/C$, but now we also get the correct no-emission
probabilities\footnote{Note that the whole procedure in principle can
  be implemented in \pytppp in a non-intrusive way, by artificially
  increasing \aem\ and implementing the reweighting and extra
  rejection in a \texttt{UserHooks} class.}. Loosely speaking we are
half the time reweighting the event to compensate for the boosting of
the emission, and half the time compensating for the corresponding
underestimate of the no-emission probability.

To see this, we again look at all possible ways of not emitting
anything between two scales, given by the modified no-emission
probability
\begin{equation}
  \tilde\Delta(t_0,t_c)=\Delta_g(t_0,t_c)\tilde\Delta_\gamma(t_0,t_c),
\end{equation}
where
\begin{equation}
  \tilde{\Delta}\gamma(t_0,t_c)=e^{-\int_{t_c}^{t_0}\tilde{\Gamma}_\gamma(t)dt}  
\end{equation}
and the product of weights from all intermediate photon emissions
which have been vetoed (with probability $1/2$, assuming we have
already taken care of the acceptance factor
$\Gamma_\gamma/\hat\Gamma_\gamma$):
\begin{eqnarray}
  \langle w\rangle\tilde\Delta(t_0,t_c)&=&
  \sum_{n=0}^{\infty}(2-\frac{2}{C})^n\tilde\Delta(t_0,t_c)
  \frac{1}{n!}\left(\int_{t_c}^{t_0}dt
    \frac{1}{2}\tilde\Gamma_\gamma(t)\right)^n\nonumber\\
  &=&\tilde\Delta(t_0,t_c)\frac{1}{n!}\left(\int_{t_c}^{t_0}dt
    \left[1-\frac{1}{C}\right]\tilde\Gamma_\gamma(t)\right)^n\nonumber\\
  &=&\tilde\Delta(t_0,t_c)\frac{1}{n!}\left(\int_{t_c}^{t_0}dt
    \left[\tilde\Gamma_\gamma(t)-\Gamma_\gamma(t)\right]\right)^n\nonumber\\
  &=& \Delta(t_0,t_c)
\end{eqnarray}
We note that we could, of course have replaced the probability one
half with any $b$, vetoing the emissions with probability $b$ and
reweighting with $(1-1/C)/b$, while reweighting with $1/((1-b)C)$ if
not vetoed, and still obtain the correct result.

We also note that while solving the same problem as was addressed in
\cite{Hoeche:2009xc}, the solution is technically different. In that
algorithm, only the standard overestimate $\hat{\Gamma}_\gamma$ is
multiplied by a factor $C$, while the acceptance of a generated
emission at scale $t$ is still done with probability
$\Gamma_\gamma(t)/\hat{\Gamma}_\gamma(t)$ and a rejected emission
instead reweights the event by a factor
\begin{equation}
  \label{eq:sherpa}
  \frac{\hat{\Gamma}_\gamma(t)-\Gamma_\gamma(t)/C}%
  {\hat{\Gamma}_\gamma(t)-\Gamma_\gamma(t)}.
\end{equation}
The end result is the same as the method presented here. In fact, if
one could choose\footnote{Note that one need to choose a $\hat\Gamma$
  which is everywhere some factor higher than $\Gamma_\gamma$ since
  otherwise the denominator in \eqref{eq:sherpa} could tend to zero,
  giving wildly fluctuating weights.} a
$\hat\Gamma_\gamma=2\Gamma_\gamma$ and let $C\to C/2$, the reweighting
of the events would be exactly the same.

While we gain in efficiency for the emissions, we will also lose in
precision for the no-emission probability due to fluctuating
weights. It is easy to calculate the variance in the weights, but it
is maybe more instructive to look at a real example.

As an illustration we let \pytppp generate standard LEP event, with
photon emission included in the shower, and we compare the default
generation with weighting the photon emission cross section with a
factor $C$. We show for different $C$ the effect of using the full
reweighting procedure (\textit{proper} weighting), but also show for
comparison the case of just using event weights with a factor
$1/C^{n_\gamma}$ (\textit{naive} weighting).

\begin{figure}
  \centering
 \epsfig{file=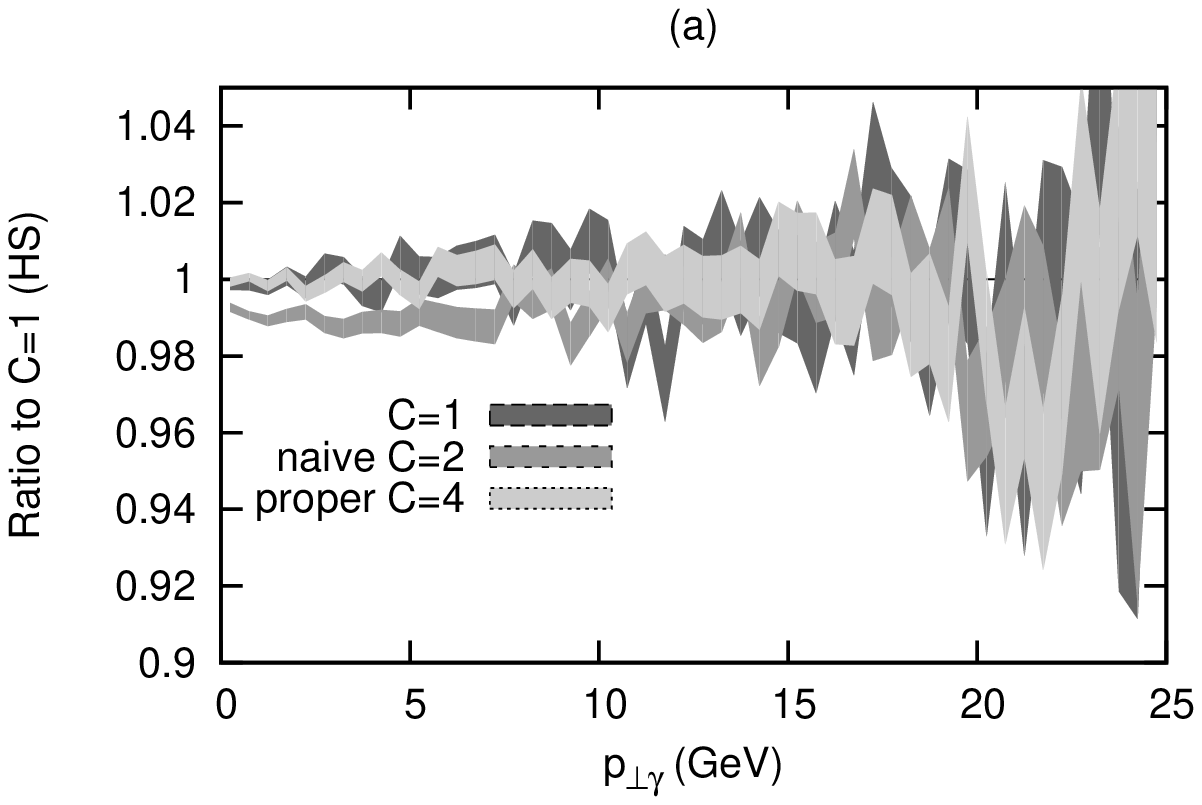,angle=-90,scale=0.5}%
 \epsfig{file=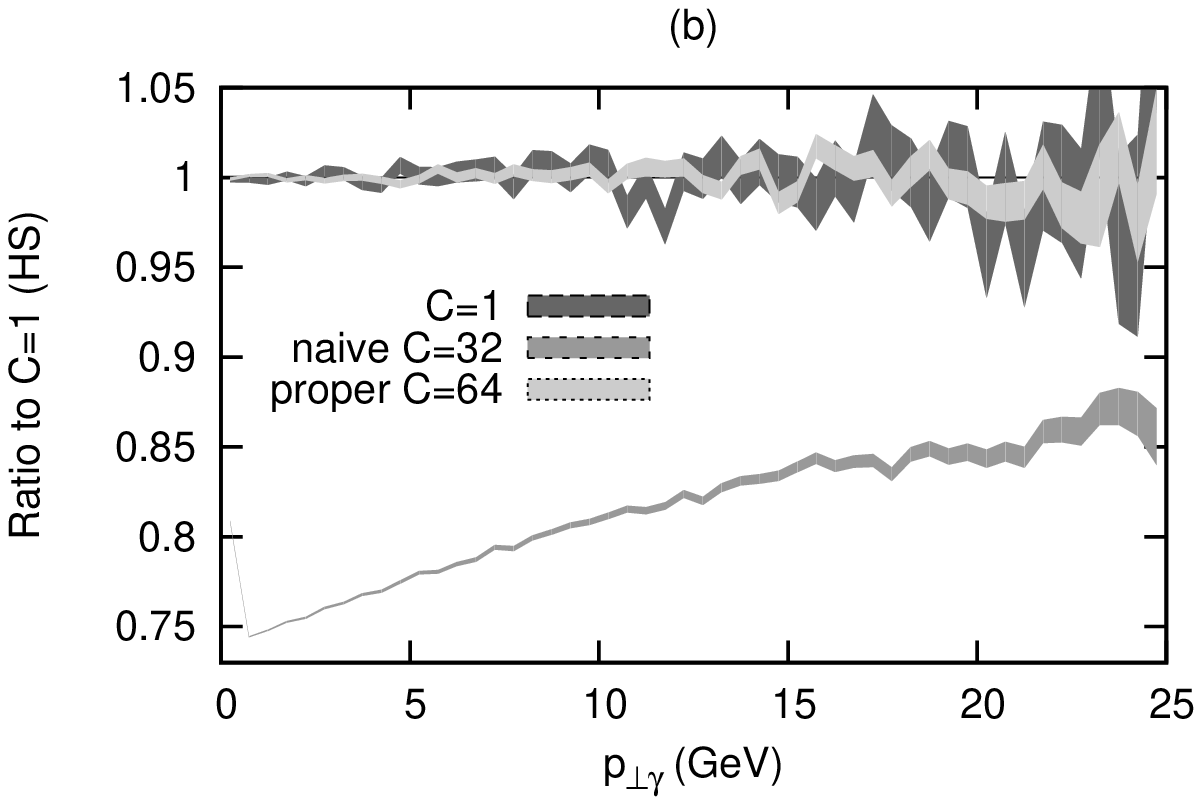,angle=-90,scale=0.5}
 \caption{\label{fig:ptgamma1}The transverse momentum (w.r.t.\ the
   thrust axis) of the most energetic photon, given as a ratio to the
   result of a high statistics run with default \pytppp result (see
   text).}
\end{figure}

In figure \ref{fig:ptgamma1}a we show the transverse momentum
distribution of the most energetic photon in an event using $C=1$
(\ie\ the default), $C=2$ for the naive, and $C=4$ for the proper
weighting\footnote{The proper reweighting has a twice as high boost
  factor, to get the same weighting of the emissions.}. The error
bands indicates the statistical error using $10^8$ events, and the
results are shown as a ratio to the result from a high statistics run
($3\cdot10^9$ events) with \pytppp. We see that the statistical error
is somewhat reduced in the reweighted samples, but we also see what
seems to be a systematic shift in the naive reweighting, due to the
mistreatment of the no-emission probability. This shift becomes very
pronounced if we increase $C$, as seen in figure \ref{fig:ptgamma1}b,
where we use $C=32$ for the naive and $C=64$ for the proper
reweighting. Here we see that the statistical errors are very much
reduced for both reweightings, but the naive procedure is basically
useless due to the large systematic shift.

\begin{figure}
  \centering
 \epsfig{file=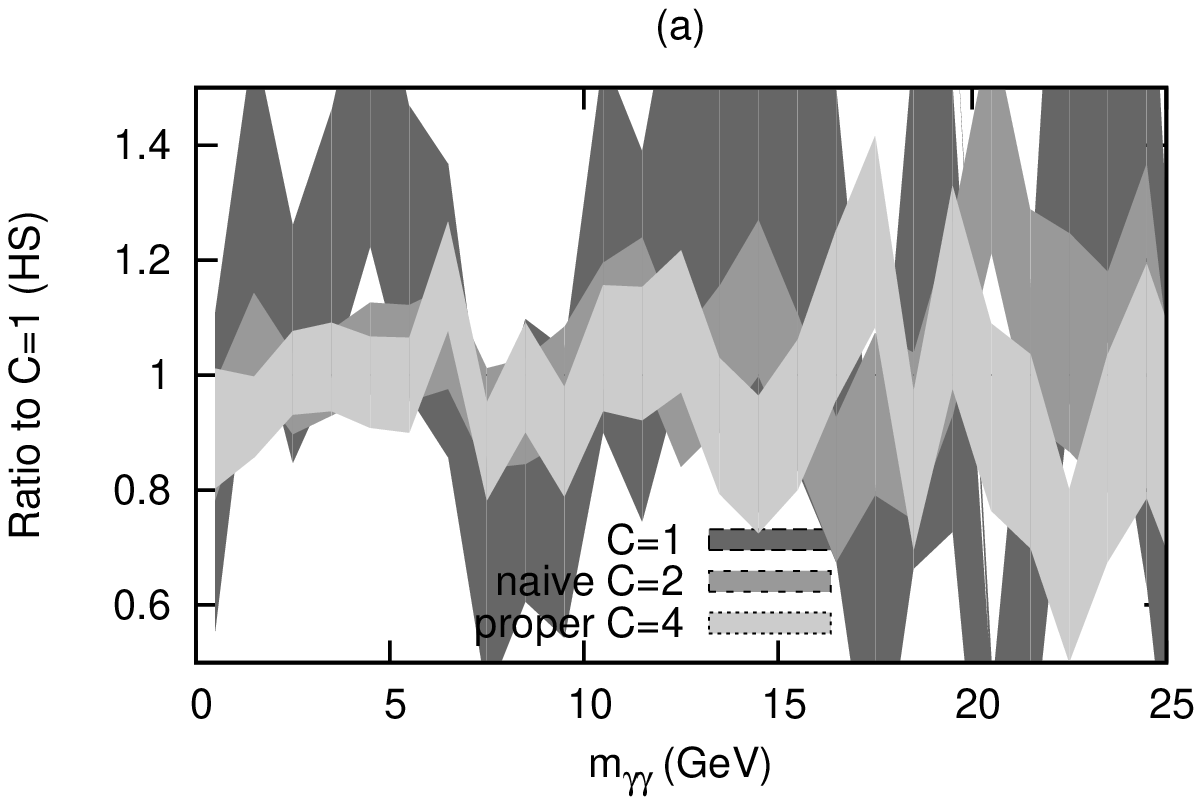,angle=-90,scale=0.5}%
 \epsfig{file=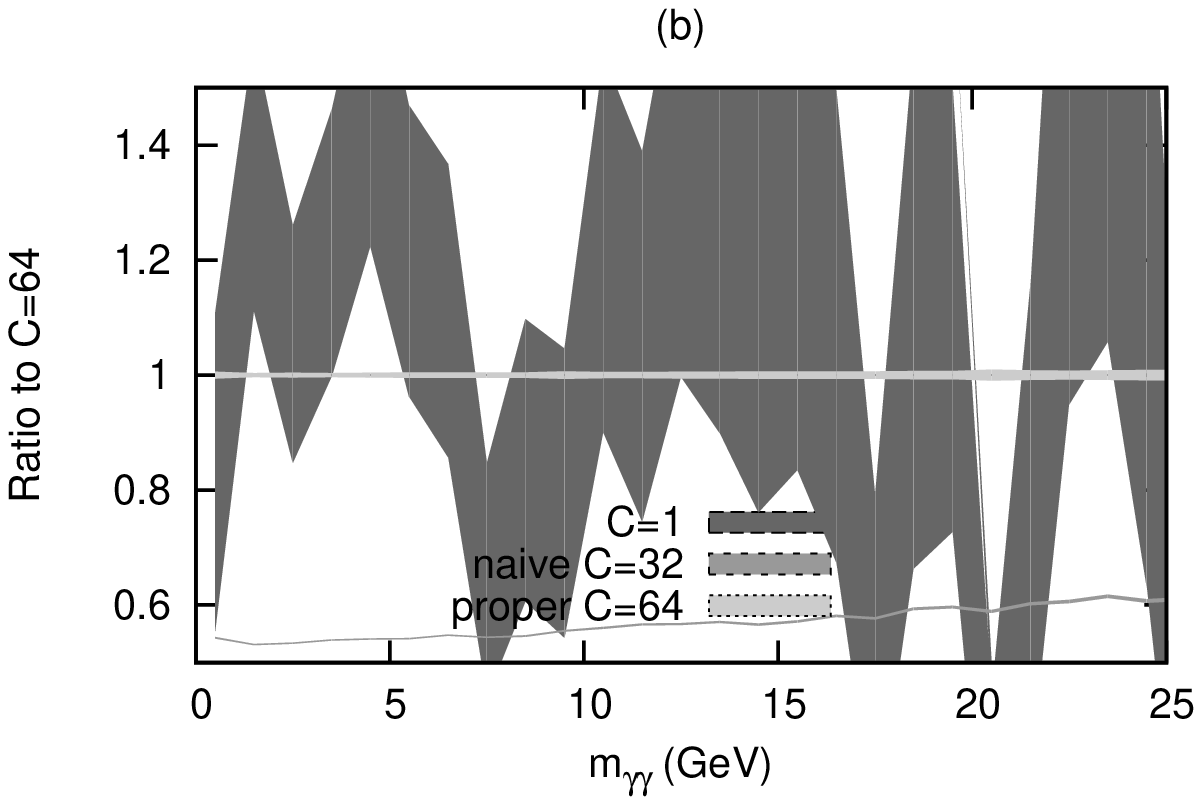,angle=-90,scale=0.5}
 \caption{\label{fig:pdigammamassR}The invariant mass of the two most
   energetic photons, given as a ratio to the result of a high
   statistics run with default \pytppp result (see text). In (b) the
   ratio is w.r.t.\ the result for the proper $C=64$ reweighting as
   even with $3\cdot10^9$ events, the statistical error from the
   default \pytppp run is too large in comparison.}
\end{figure}

If we require two photons in each event, the gain from the reweighting
becomes more obvious. In figure \ref{fig:pdigammamassR} we show the
distribution in invariant mass of the two most energetic photons in an
event. Here the gain in statistics is significant also for the case of
modest boost factors (a), and for the large boost factors in (b) the
gain in statistics is enormous, while, again, the naive reweighting
suffers a large systematic shift.

\begin{figure}
  \centering
 \epsfig{file=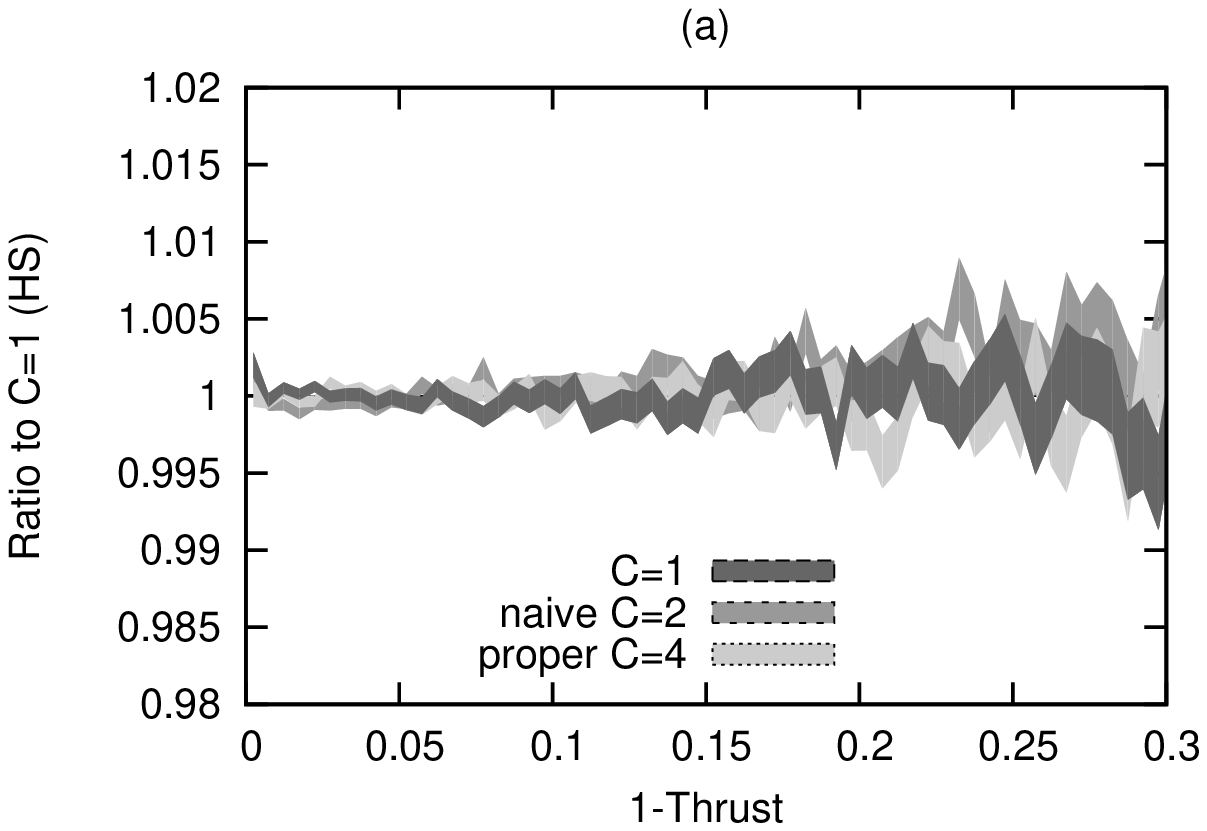,angle=-90,scale=0.5}%
 \epsfig{file=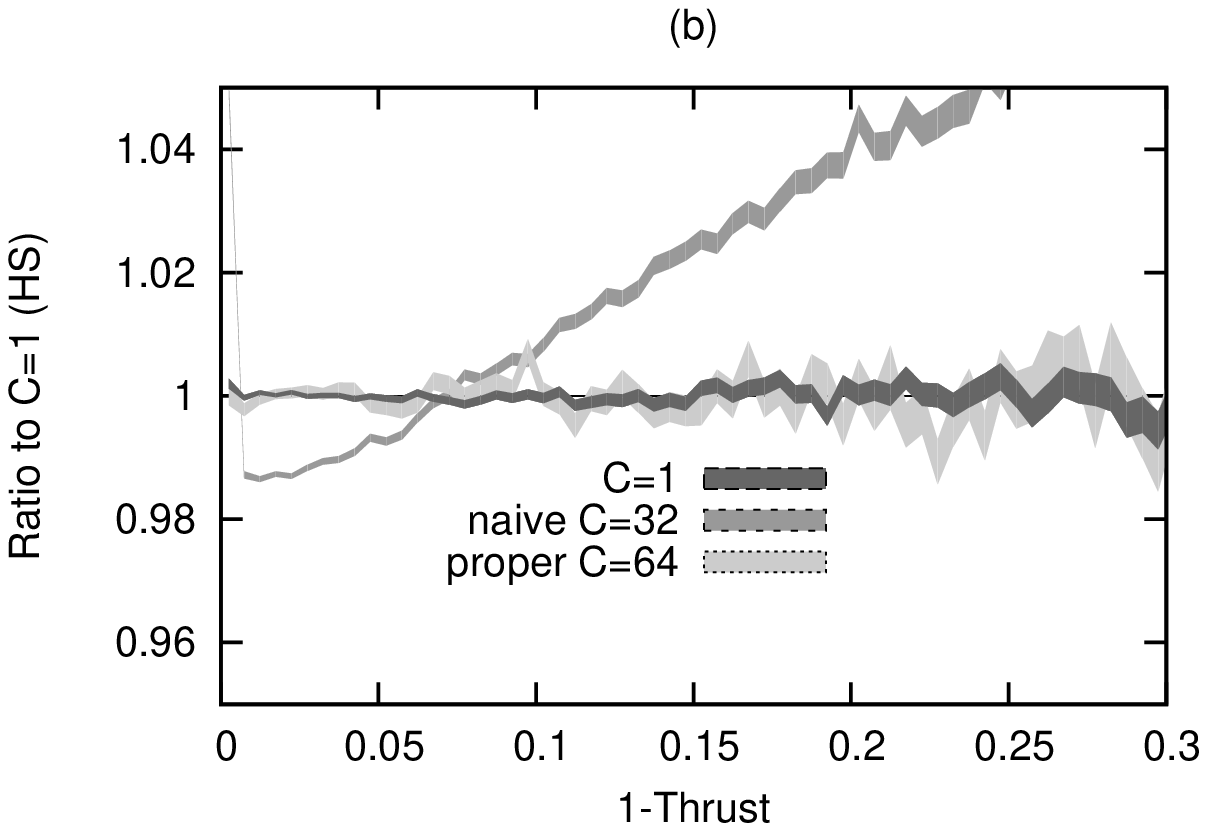,angle=-90,scale=0.5}
 \caption{\label{fig:thrust}The inclusive thrust distribution, given
   as a ratio to the result of a high statistics run with default
   \pytppp result (see text).}
\end{figure}

To isolate the effect on the no-emission probability, figure
\ref{fig:thrust} shows the inclusive thrust distribution for the same
runs as before. Here we see that especially with forceful proper
reweighting the statistical error is increased because of the
fluctuating weights, and we see again that the naive reweighting will
give a systematic shift due to the mistreatment of the no-emission
probability.

So far we have implicitly assumed that $C>1$, since we motivated the
whole procedure by the desire to increase the number of rare
splittings. Note, however, that the proof of the procedure does not at
all depend on the size of $C$. In fact it can even in some cases be
taken negative.

Consider a case where there are negative contributions to the total
splitting probability. One of the most simple cases is the emission of
a second gluon in a final-state dipole shower in \tee-annihilation
into jets. Once a gluon has been radiated from the original $\q\qbar$
pair, it can be shown that the distribution of a second gluon is well
described by independent emissions from two dipoles, one between the
quark and the first gluon and one between the gluon and the
anti-quark. However, examining the $\ee\to\q gg\qbar$ matrix element
one finds that there is a colour-suppressed negative contribution
corresponding to emissions from the dipole between the $\q$ and
$\qbar$. This contribution is normally ignored completely in parton
showers, mainly because it is difficult to handle in a probabilistic
way in the SVA. It may even result in a no-emission probability above
unity.

In the reweighting scheme introduced here we can easily include the
negative contribution to the splitting functions, and apply a boost
factor of $C=-1$ for the $\q\qbar$-dipole. If a gluon emission is
generated from such a dipole, it is then either accepted and the
event is given a negative weight, or it is rejected, in which case the
event weight is multiplied by a factor four. We note that in this way
it is in principle conceivable to implement a parton shower which
includes all possible interference effects. We will, of course, have
even larger issues with statistics, compared to the photon emission
case above, as we now have potentially large weights that must cancel
each other, but this procedure could still be an interesting
alternative to the ones presented in \cite{Platzer:2011dq} and
\cite{Hoeche:2011fd} (an extension of \cite{Hoeche:2009xc} to negative
weights).

\section{Conclusions}
\label{sec:conclusions}

This article does not claim to present innovative new physics
results. Rather it presents a number of methods collected by the
author during a couple of decades working with parton showers in
general and with the Sudakov veto algorithm in particular. They are
presented here in the hope that they may come in handy for the
community now that more and more efforts are put into the merging and
matching parton showers with matrix element. Especially in the case of
matching with next-to-leading order matrix elements (and beyond), a
thorough understanding of how parton showers work and knowledge of how
to manipulate them is necessary, and these kinds of methods may become
increasingly important.

\bibliographystyle{utcaps}
\bibliography{refs}

\end{document}